\documentclass[a4paper,preprint,14pt]{revtex4}
\usepackage{graphicx}
\usepackage{courier}
\usepackage{amsmath}
\usepackage{prettyref}

\graphicspath{{figures/}}

\newcommand{\cm}{cm$^{-1}$}
\newcommand{\braket}[3]{\left< #1 | #2 | #3 \right>}
\begin{document}
\title{Rotational cooling efficiency upon molecular ionization: the
case of Li$_2(a^3\Sigma_u^+)$ and Li$_2^+(X^2\Sigma_g^+)$ interacting
with $^4$He}

\author{M. Wernli}
\affiliation{Dept. of Chemistry and CNISM, University of Rome "la
Sapienza", Ple A. Moro 5, 00185 Rome, Italy}

\author{E. Bodo}
\affiliation{Dept. of Chemistry and CNISM, University of Rome "la
Sapienza", Ple A. Moro 5, 00185 Rome, Italy}

\author{F.~A. Gianturco}
\affiliation{Dept. of Chemistry and CNISM, University of Rome "la
Sapienza", Ple A. Moro 5, 00185 Rome, Italy}
\
\email{fa.gianturco@caspur.it}

\date{\today}

%
%
\begin{abstract}The low-temperature (up to about 100K) collisional
(de)excitation cross sections are computed using the full
coupled-channel (CC) quantum dynamics for both Li$_2$ and Li$_2^+$
molecular targets in collision with $^4$He. The interaction forces are
obtained from fairly accurate {\it ab initio} calculations and the
corresponding pseudo-rates are also computed. The results show
surprising similarities between sizes of inelastic flux distributions
within final states in both systems and the findings are connected
with the structural change in the molecular rotor features when the
neutral species is replaced by its ionic counterpart.
\end{abstract}
%
%
%

\maketitle

\section{\label{intro}Introduction}
Lithium-bearing molecules have been a source of research interest for
several years, since they may have played an important role as coolers
(through rotational transitions) in the young universe (see
\cite{lithium_rep} for a detailed review of the topic). Their
usefulness in the study of molecular dynamics at ultralow temperatures
has also been demonstrated recently in relation with their possible
role in molecular formation under Bose-Einstein condensation
conditions in magneto-optical traps
\citep{launay05,launay07,launayPRL,grimm05}. Two types of lithium
dimers are studied here in relation to their collisions with $^4$He
atoms: Li$_2(a^3\Sigma_u^+)$ and Li$_2^+(X^2\Sigma_g^+)$, hereafter
simply called ``neutral'' and ``ion''. Although they are not easily
observable because of the absence of a permanent electric dipole
moment, their comparative study is interesting in several respects,
e.g. the change upon ionization within the full manifold of internal
state transitions of the collisional behavior and the importance of
the quadrupolar ($\Delta J=2$) transitions. Furthermore, although the
lithium dimer could be considered the second simplest homonuclear
molecule after H$_2$, only a few recent studies (see \cite{minaev05}
and references therein) have begun to give information on its
chemistry, spectroscopy and collisional properties. Hence, still a lot
of work has to be done to fully understand this arguably simple
case. The scope of the present paper is thus to compare the dependence
of collisional quantities, i.e. cross sections and rates, on changing
the electronic state of the dimer target.\\

The paper is organized as follows: Section \ref{sec:methods} outlines
the computation methods and the numerical algorithm employed in this
study. Section \ref{sec:results} reports the results of our scattering
calculations, with an analysis of the similarities/differences between
the two title systems. Section \ref{sec:conclusion} summarizes our
conclusions.

\section{\label{sec:methods}The computational procedure}
\subsection{\label{sec:2}Analytic fitting of the potential energy surfaces}
The two potential energy surfaces (PES) used in the present work have
already been computed from {\it ab initio} calculations carried out at
the MP4 {\it ab initio} level \cite{bodosurf}. Both systems are
treated here as rigid rotors, with bond lengths fixed at their
equilibrium values : 4.175\AA\ for the neutral, 3.11\AA\ for the
ion. We considered in all calculations only the dominant isotope of
Li, $^7$Li. Accordingly, we used in the dynamical calculations
rotational constant $B$ values of 0.2758 and 0.4971 \cm,
respectively. \\The difference in bond lengths when ionization takes
place appears to be a peculiar property of the present system and is
by no means a general feature of diatomic targets undergoing
ionization, as shown by the data of table \ref{table:species}.

\begin{table}
  \begin{tabular}{|c|c|c|c||c|c|c|c|c||c|c|c|}
    \hline
    mol. & state & r$_e$ [a$_0$] & B [cm$^{-1}$] \\
    \hline
    Li$_2$ & $X^1\Sigma_g^+$ \cite{lidata}& 5.10 & 0.660 \\
    \hline
    Li$_2^+$ & $X^2\Sigma_g^+$ \cite{lidata}& 5.88 (15\%) & 0.4971 \\
    \hline
    Li$_2$ & $a^3\Sigma_u^+$ \cite{lidata}& 7.89 (55\%) & 0.2758 \\
    \hline
    \hline
    Na$_2$ & $X^1\Sigma_g^+$ \cite{ho} & 5.82 & - \\
    \hline
    Na$_2^+$ & $X^2\Sigma_g^+$ \cite{patil}& 6.8 (17\%)& -\\
    \hline
    Na$_2$ & $a^3\Sigma_u^+$ \cite{ivanov}& 9.76 (68\%) & - \\
    \hline
    \hline
    K$_2$  & $X^1\Sigma_g^+$ \cite{heinze} & 7.41 & - \\
    \hline
    K$_2^+$  & $X^2\Sigma_g^+$ \cite{patil}& 8.3 (12\%) & - \\
    \hline
    K$_2$  & $a^3\Sigma_u^+$ \cite{jong}& 10.91 (47\%) & - \\
    \hline
    \hline
    H$_2$ & X$^1\Sigma$ & 1.40 & 60.85 \\
    \hline
    H$_2^+$ & X$^2\Sigma_g^+$ & 1.98 & 30.20 \\
    \hline
    \hline
    O$_2$ & X$^3\Sigma_g^-$ & 2.29 & 1.438 \\
    \hline
    O$_2^+$ & X$^2\Pi_g$ & 2.12 & 1.691 \\
    \hline
    \hline
    Ne$_2$ & X$^1\Sigma_g^+$ & 5.86 & 0.17 \\
    \hline
    Ne$_2^+$ & X$^2\Sigma_u^+$ & 3.31 & 0.55 \\
    \hline
  \end{tabular}
  \caption{Comparison of molecular parameters for homonuclear neutral
    and ionic molecules in their electronic ground and first excited
    states. Figures in parenthesis give the percentage lengthening of
    the bond with respect to the $^1\Sigma$ reference state. Numbers,
    if not directly referenced, are taken from the NIST diatomic
    database \cite{nist}.}
  \label{table:species}  
\end{table}

It is interesting to understand why the bond distances become longer
in the case of the ionic doublets and even more so for the neutral
triplets. The changes of core orbitals along the alkali metal sequence
are balanced by the increase in atomic numbers that create more
attractive Coulomb wells around the nuclei. Thus, the outer electrons
(one or two) play a very similar role in all three systems, going from
Li dimers to K dimers. The crucial difference thus comes from the
Pauli repulsion occurring between the outer electrons of the
$^3\Sigma$ case (they have aligned spins), which is even stronger than
the additional repulsive contribution among core electrons and the
single outer one that is a consequence of the reduction of the
screening of nuclear charges caused by the ionization process. Such an
effect is not observed for the non-alkali dimers reported in table
\ref{table:species}, where we observe always bond contraction after
molecular ionization processes, except for the H$_2$ dimer which, with
only two bound electrons, is another system which follows the alkali
metal behavior. In conclusion, the two title systems show very marked
bond lengthening both upon ionization and on spin stretching.  \\

To solve the close coupling equations, it is necessary to generate the
matrix elements of the coupling potential between the basis of
asymptotic functions. Since the latter are given by Legendre
polynomials, we fit the potential as follows

\begin{equation}
  V(R,\theta|r_{eq})= \sum_\lambda V_\lambda(R|r_{eq})
  P_\lambda(\cos(\theta))
  \label{eq:fit}
\end{equation}

where $R$ is the intermolecular distance (distance from the center of
mass of the dimer to the atom), $r_{eq}$ is the diatomic rigid rotor
bond distance, $\theta$ is the angle between the dimer and the
intermolecular vector, $P_\lambda$ are the Legendre polynomials, and
$V_\lambda$ are the radial coefficients. The latter are the potential
coupling coefficients which shall be employed in the scattering
equations. We thus need to evaluate them at any $R$, with $\lambda$
large enough for the expansion to reach a preselected precision. In
practice, we solve it over a discrete radial grid, and then
interpolate the $V_\lambda$ with cubic splines, further extrapolating
them with exponentials at short range and a two-term inverse power law
at long range.\\

Both systems, although showing different potentials, are strongly
anisotropic in the short range region: for some chosen value of $R$,
the potential can therefore be for different angles either strongly
attractive or repulsive by several thousands \cm. This feature makes
it numerically difficult when trying to generate the radial
coefficients. A method has been applied (e.g. see \cite{HC3N}) which
permits to circumvent this difficulty: at a given $R$ value, we first
truncate the potential up to a few thousand \cm and then apply a
smoothing function to this truncated potential to avoid the Gibbs
oscillations that would inevitably come when fitting directly the
truncated potential. Finally, we fit this functional of the potential
with a weighting strategy, giving more relevance (thus higher fitting
precision) to the low-energy parts of the potential in comparison with
the more repulsive regions. Thus, we optimized the fitting parameters
for the low-energy dynamics which we intend to study by finally
getting potential fits with a precision of better than 2 \cm\ for
$V\lesssim 800$ \cm.

\subsection{\label{sec:dynamics}The quantum dynamics}
We briefly recall here the equations of the close-coupling formalism
we have employed. Using the center of mass frame, the time-independent
Schr\"odinger equation writes
\begin{equation}
  \left(T_r + T_R + v_{mol}(r) + V_I(r,R,\theta) - E\right)
  \Psi^{JM}(\vec{R},\vec{r}) = 0
  \label{FC1}
\end{equation}
where
\begin{eqnarray}
  T_r &=& -\frac{1}{2m} \nabla_r^2 \quad{\rm and}\\
  T_R &=& -\frac{1}{2\mu} \nabla_R^2
\end{eqnarray}
where $m$ is the reduced mass of the diatom and $\mu$ that of the
complex. E is the total energy. The interatomic distance of the diatom
is denoted $r$, $R$ is the distance between the colliding atom and the
diatom center of mass and $\theta$ is the angle between $\vec{R}$ and
$\vec{r}$. The $v_{mol}$ term is the potential of the isolated diatom
and $V_I(r,R,\theta)$ is the interaction potential,
eq. (\ref{eq:fit}). To solve equation (\ref{FC1}), the
$\Psi^{JM}(\vec{R},\vec{r})$ is expanded on a basis of asymptotic
eigenfunctions of the isolated partners, which are treated here as
rigid rotor targets ($\overline r =\overline r_{eq}$ for each of them)
and therefore $\overline r_{eq}$ disappears as an explicit variable of
the present problem.
\begin{equation}
  \Psi^{JM}(\vec{R},\hat{r}_{eq})=\frac{1}{R}\sum_n C_n(R)
  \phi_n(\hat{R},\hat{r}_{eq})
  \label{FC2}
\end{equation}
where the channel function for channel $n\equiv (j l\,;J M)$ is given
by
\begin{equation}
  \phi_n(\hat{R},\hat{r}_{eq})= \sum_{m_j,m_l}(j,l,J|m_j,m_l,M)
  Y^j_{m_j}(\hat{r}_{eq}) Y^l_{m_l}(\hat{R})
  \label{FC3}
\end{equation}
The quantum number for rotation is denoted by $j$ and $l$ is the
orbital angular momentum of the atom with respect to the diatom. $J$
is the total angular momentum ($\vec{J}=\vec{j}+\vec{l}$), $M$ is the
projection of $J$ on the laboratory frame fixed $z$ axis, and
$(j,l,J|m_j,m_l,M)$ is a Clebsch-Gordan coefficient. Solving the
present problem is thus equivalent to determining the expansion
coefficients $C_n(R)$. Multiplying the l.h.s. of eq. (\ref{FC1}) by
$\phi_i$ and integrating over $\hat{R}$ and $\hat{r}$, then using
eq. (\ref{FC2}) and (\ref{FC3}), we find
\begin{equation}
  \left(\frac{d^2}{dR^2} -\frac{l_i(l_i+1)}{R^2}+2\mu E_i \right)
  C_i(R) = 2\mu \sum_n C_n(R) \braket{\phi_i}{V_I}{\phi_n}
  \label{FC5}
\end{equation}
where $E_i=E-\epsilon_j$ is the initial kinetic energy (the collision
energy) and $l_i$ is the angular orbital momentum in the $i$th
channel. $\epsilon_j=Bj(j+1)$ is the rotational energy of the
target. We obtained a second-order differential equation to be solved
for each $i$, thus a set of equations for the $C_i(R)$ coefficients,
called coupled channel (CC) equations.

The radial coefficients $V_\lambda(R)$ hence appear in the sum of
terms on the r.h.s. of eq. (\ref{FC5}),as the weighting radial terms
of the potential times the angular coupling terms generated by the
potential anisotropy between rotational asymptotic channels
$\braket{\phi_i}{P_\lambda}{\phi_n}$. We thus know that the angular
dependence of the interaction applies, during collisions, a torque to
the rotating target which acts over the radial range of action of each
$V_\lambda(R)$ coefficient.\\

To solve the CC equations we used the code developed in our group,
where the propagator was given by a log-derivative algorithm at short
range and by the modified variable-phase propagator at long range, as
discussed by \cite{var_ph}.

The initial tests for the neutral showed that the inelastic cross
sections would exhibit very few and fairly small resonance
features. Accordingly, we chose a rather sparse energy grid
corresponding to a minimum of 50 energies for the highest initial $j$,
this number increasing with decreasing initial $j$. The energies were
chosen to be mainly distributed around the expected isolated resonance
energies, in order to obtain a good description of these features.\\
The propagators parameters were accurately tested at two
representative energies and the integration was thus carried out using
the Log-Derivative propagator between 2 and 30 \AA\ (in 500 steps),
and our Modified Variable Phase propagator between 30 and 200 \AA.\\
The rotational basis chosen covered a range of more than 100 \cm\ for
the closed channels at all energies and it proved to be necessary to
compute partial cross-sections up to a total angular momentum of
35$\hbar$ in order to get a satisfactory convergence (around 5\%), at
the higher rotational transitions.\\ Our results were further tested
at a few energies using an entirely different code by Hutson and Green
(see ref. \cite{molscat}). We found an excellent agreement between the
two codes, with differences remaining always under 1\%.\\

When computing rotational transitions for the ion the larger potential
depth induces a much richer resonance structure. We consequently
adopted a more dense energy grid and a larger rotational basis. The
minimum number of energy points for all transitions is here about 100
in the energy range of 1-100 \cm\ and we also computed a few points up
to 600 \cm\ to further ensure numerical convergence of our rate
coefficients (see below for additional details). For the ion, we used
more or less the same propagation parameters as for the neutral, while
just switching from one propagator to the other at an earlier distance
of 15 \AA. We also used at all energies a rotational basis equivalent
to at least a range of 300 \cm\ spanned by the energies of the closed
channels. The maximum total angular momentum needed below 100 \cm\ was
of $\sim$45$\hbar$.

For both systems the detailed balance on the cross sections gave an
excellent agreement at all energies, the largest error coming at low
energies, with differences around 5\%. This permits us to state that
our final cross sections are numerically converged within that error
value.

Since the main scope of this paper is the comparison of rotational
(de)excitation behavior of the two systems, the spin coupling effects
(spin-spin and spin-rotation) are neglected and both systems are
treated as pseudo-$^1\Sigma$ targets. This approximation is fully
justified at our energies, since spin coupling constants are small for
both systems. According to Kurls's formula (e.g. see ref. \cite{kurl})
and using the data provided to us by E. Yurtsever (private
communication) obtained via the Gaussian code (e.g. see ref.
\cite{gaussian}), we find that the spin-rotation constants are
respectively $4.8\ 10^{-5}$ and $4.3\ 10^{-5}$ \cm\ for the neutral
and the ion. We have nonetheless performed a few numerical tests with
the correct coupling calculations at several energies between 1 and 30
\cm, and for both systems we found that the inclusion of the
spin-rotation coupling has only a small effect at the energies we
considered. If we sum over final spin states, the value of a given
rotational transition, in fact, does not depend on the initial choice
by more than 5\%. Furthermore, the difference between the summed cross
section and its value from the pseudo-$^1\Sigma$ calculation is less
than $10$\%. At all the energies of this range, moreover, the systems
preferentially stay in their original spin state, this preference
varying from a factor of $\sim 1.5$ to more than 10. These results
confirm the validity of the pseudo-singlet approximation employed in
our extensive calculations reported below.

\section{\label{sec:results}Results and discussion}
\begin{figure}
  \begin{center}
    \begin{tabular}{cc}
      \includegraphics[width=0.5\textwidth]{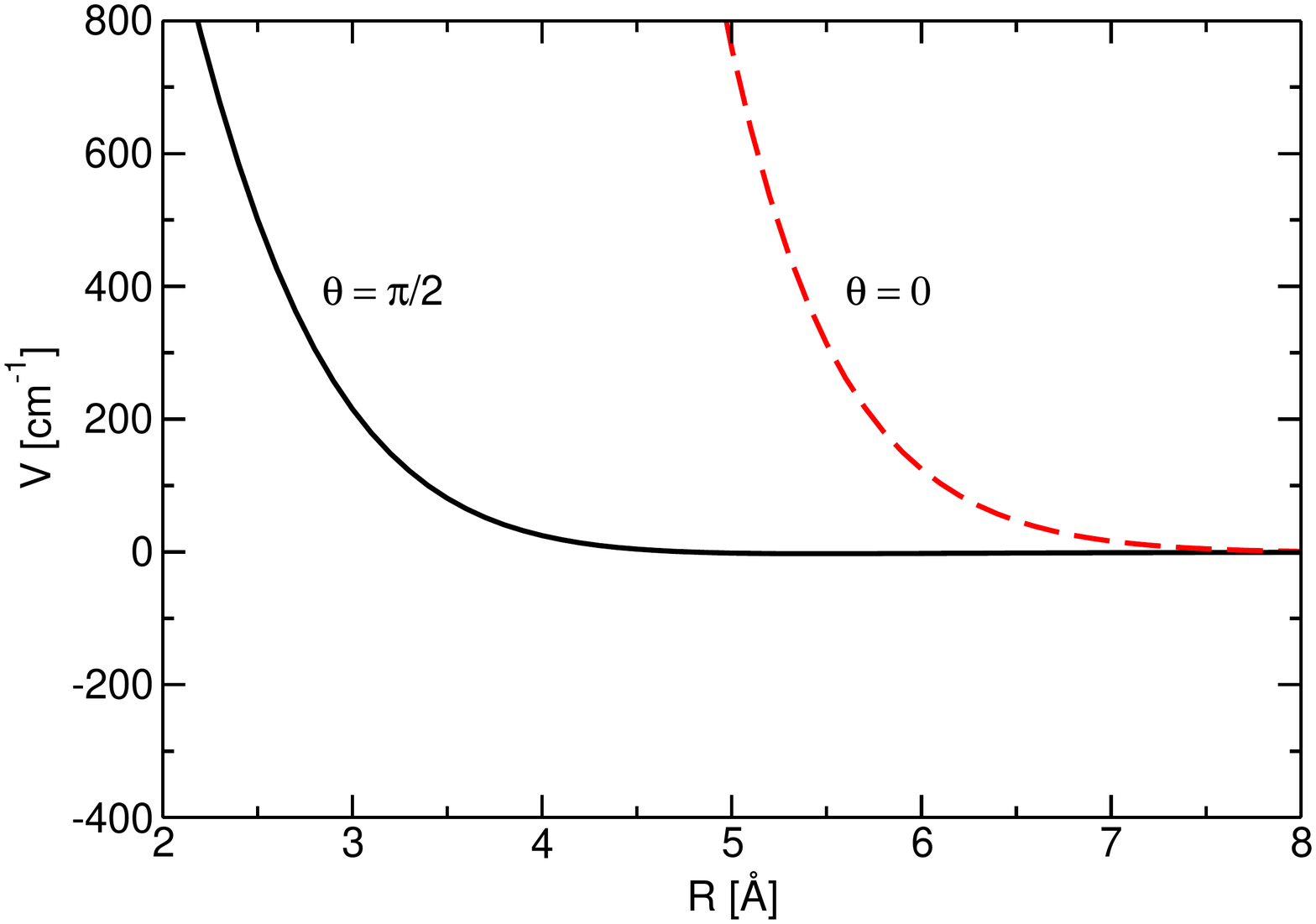} &
      \includegraphics[width=0.5\textwidth]{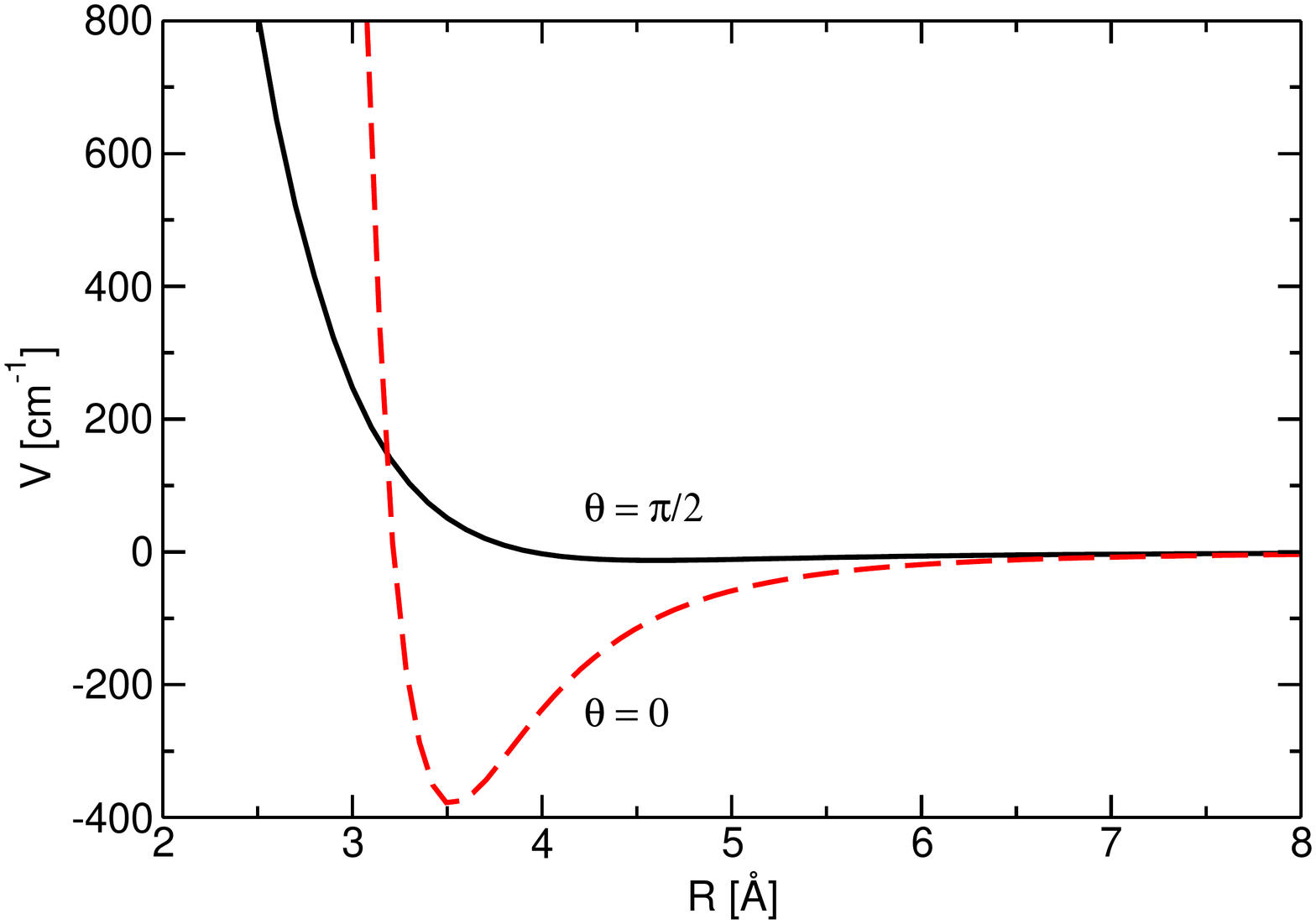}
    \end{tabular}
    \caption{Potential energy curves for neutral and ionic Li$_2$
    interacting with $^4$He, as a function of intermolecular distance
    for the collinear ($\theta=0$) and perpendicular ($\theta=\pi/2$)
    relative orientations. Left panel: Li$_2$; right panel:
    Li$_2^+$. The global minimum of Li$_2$-He interaction is -2.3
    \cm.}
    \label{fig:pot}
   \end{center}
\end{figure}

Figure \ref{fig:pot} shows the potential energy curves resulting from
our fit for both systems and reports their minimum orientations. The
two surfaces markedly differ in several points: the potential well
depth is more than a hundred times deeper in the case of the ion and
the repulsive walls at short range do not have the same slope, neither
the same location. But two facts are nonetheless common to both
molecular partners: (i) the presence of an attractive interaction in
the medium to long range region and a strongly repulsive wall when
approaching at short range each molecule and (ii) the presence of only
two dominant multipolar potential terms at long range, i.e. $V_0(R)$
and $V_2(R)$. For the ion, they correspond respectively to the
charge-induced dipole and charge-induced quadrupole interactions. For
the neutral partner, we have instead the isotropic and anisotropic
part of the dispersion interaction, respectively. Both terms for the
neutral are smaller than for the ion. Moreover, the $V_0$ vanishes
more rapidly for the neutral. In conclusion, the neutral triplet state
exhibits ``softer'' repulsive regions than the more compact ionic
doublet.\\

\begin{figure}
      \includegraphics[width=1\textwidth]{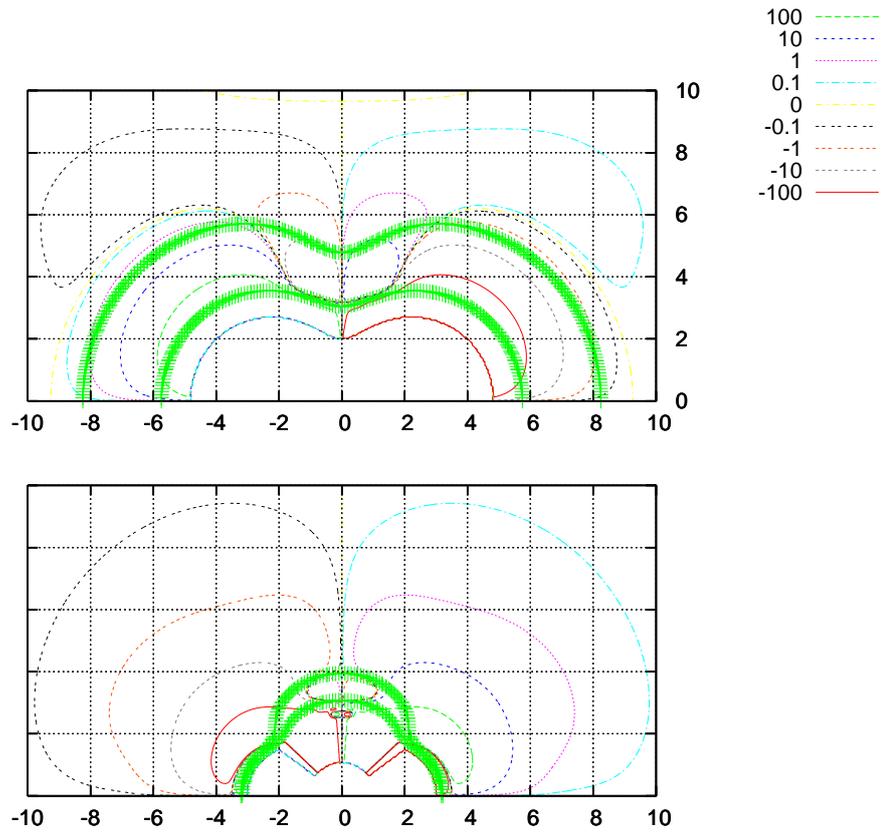}
    \caption{Torques ($\partial V/\partial \theta$) isolines for the
    Li$_2$ (upper panel) and Li$_2^+$ (lower panel) as a function of
    the cartesian coordinates $x$ and $y$, in \AA. The zero potential
    isolines those at 200 \cm\ are given by thicker lines. The energy
    isolines scale is logarithmic: -100, -10, -1, -0.1, 0, 0.1, 1, 10,
    100 \cm.}
    \label{fig:torques}
\end{figure}

Figure \ref{fig:torques} shows the computed ``potential torques''
($\partial V/\partial \theta$) for both systems, using the same unit
scale and as a function of the $x$ and $y$ cartesian coordinates. As
seen before with the potential curves, the computed torques have very
different ranges of action, although they have in common that the most
efficient angular coupling for both systems is in the range of
$\theta=30-40^o$. From the shape of the angular torques, combined with
the features of figure \ref{fig:pot}, we see that the helium atom can
get closer to the molecular partner in the case of the neutral (its
repulsive wall is less steep) than in the ionic case, so that the
overall torques sampled at a given collision energy are of the same
order of magnitude for both systems in the sense that the weaker
torque applied by the incoming He atom to the neutral partner has a
much larger range of action during collision than in the case of the
stronger torque applied to the ionic partner. If we combine this
finding with the reduced energy gaps between rotor states of the
triplet when is compared with the ion, we see that the neutral
interaction, albeit weaker, becomes just as efficient in exciting
rotations as the corresponding ionic target.\\

\begin{figure}
  \resizebox{15cm}{!}{\includegraphics{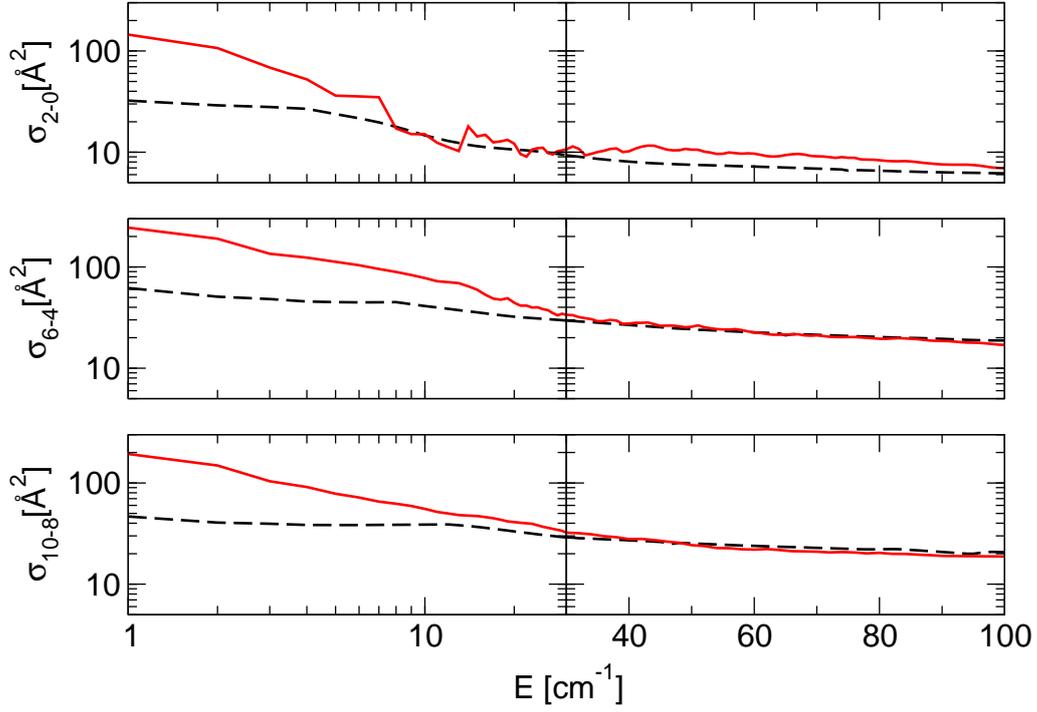}}
  \caption{Computed rotational deexcitation cross sections for both
    systems as a function of collision energy. The calculations for
    the ion are given by solid lines, while those for the neutral are
    given by dashes. The three transitions represented are, from top
    to bottom, those involving the following levels: 2-0, 6-4,
    10-8. The energy scale is logarithmic between 1 and 30 \cm, then
    linear at higher energy.  } 
  \label{fig:sections}
\end{figure}

Figure \ref{fig:sections} shows some illustrative results obtained for
the cross sections. The most surprising finding is that at collision
energies of a few \cm\ above threshold the deexcitation cross sections
are of the same order of magnitude for both systems. One would have
expected that the much deeper potential well depth and the greater
strength of the long-range forces would cause larger cross sections
for the ion. On the other hand, the foregoing discussion on the range
of action of the rotational torques acting during collision provides a
structural explanation for the size similarities between cross
sections. We should also note that ref. \cite{HC3N} has already shown
that for large enough bond distances the collisional behavior is
dominated by the geometry of the target molecule and classical
calculations provide good agreement with quantum
results. Consequently, we should expect that a classical approach to
rotational cooling may also work reasonably well for the two title
systems of the present work.\\ One should also note here that the
oscillatory structures are much richer in the case of the ion,
occurring up to 50 \cm\ above threshold. It is hard to decide whether
these structures are due to resonant features or to background
interference effects without a proper analysis of the corresponding
S-matrix elements. As we consider such a study, because of the absence
of experimental data, outside the scope of this paper, we are not
discussing these features any more. For the neutral, on the other
hand, only one small feature in the cross sections appears, associated
with the opening of the first rotational channel. Apart from such
low-energy findings, the behavior of all cross sections is largely
featureless as the energy increases. In both cases, rotational
deexcitation is a more favorable process when starting from higher $j$
values. Hence, rotational excitation is expected to be easier from low
$j$ initial states. As noted before, the largest difference between
the two systems comes at low energies, where the strong increase of
cross sections is much more marked in the case of the ion as the
energy decreases: clearly, at low energies, the systems are more
sensitive to the outer potential details like well depth and long
range forces, the latter dominating the threshold behavior in the
ionic system.

If we further define a quantity we shall call the pseudo-rates $K$ as
$K_{j\to j'}(E)= \sigma_{j\to j'} v(E)$, where $v(E)$ is the velocity
associated to the initial collision energy E : $v=\sqrt{2E/\mu}$\ , we
note that these quantities are generally a good approximation of true
Boltzmann rates when the cross sections are nearly featureless and
smoothly vary with temperature, given as $E=kT$ in the
pseudo-rates. As discussed for the data shown by figure
\ref{fig:sections}, this is what occurs in the present calculations.

\begin{figure}
  \begin{center}
    \begin{tabular}{cc}
      \includegraphics[width=0.5\textwidth]{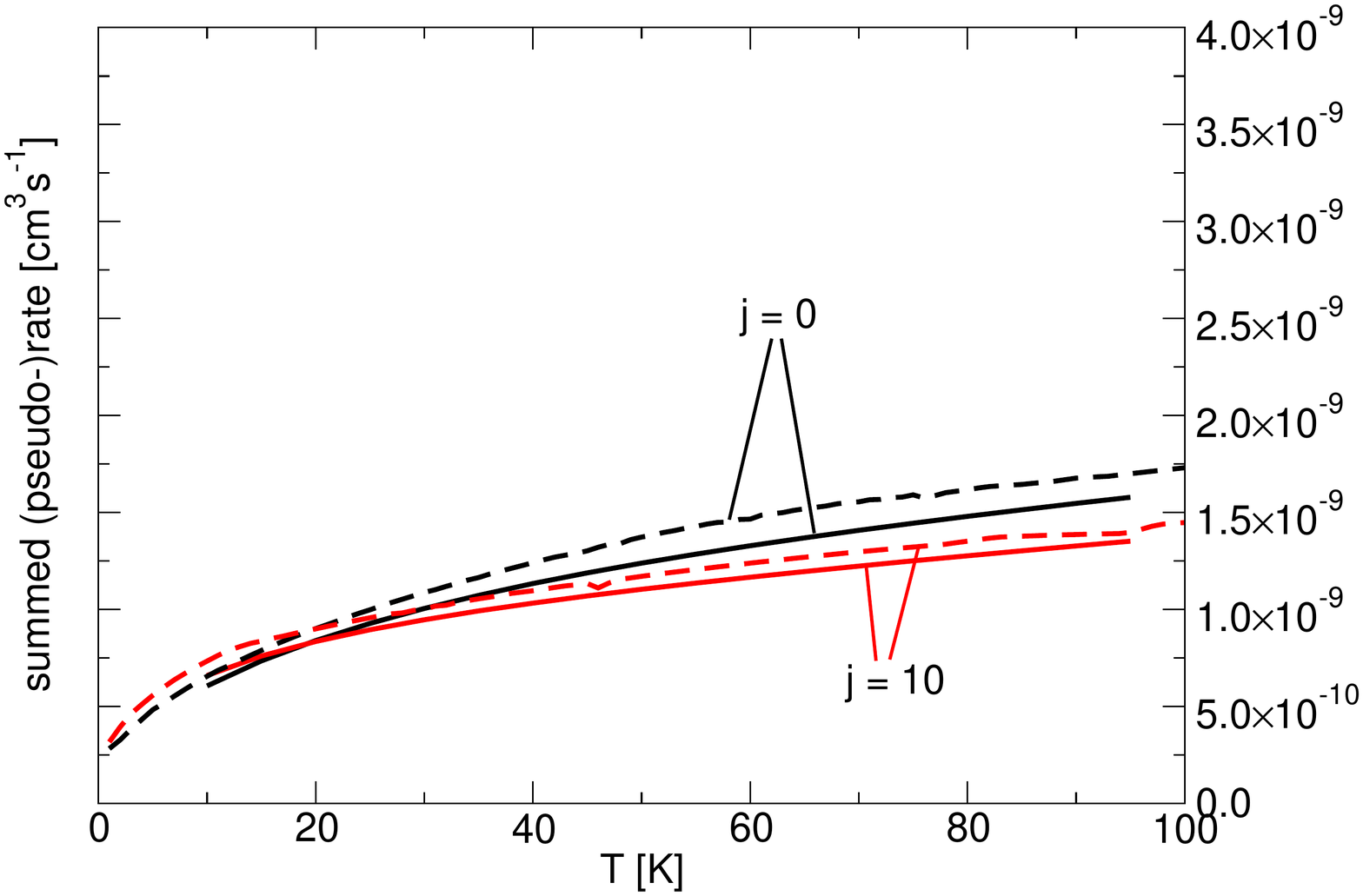}
      &
      \includegraphics[width=0.5\textwidth]{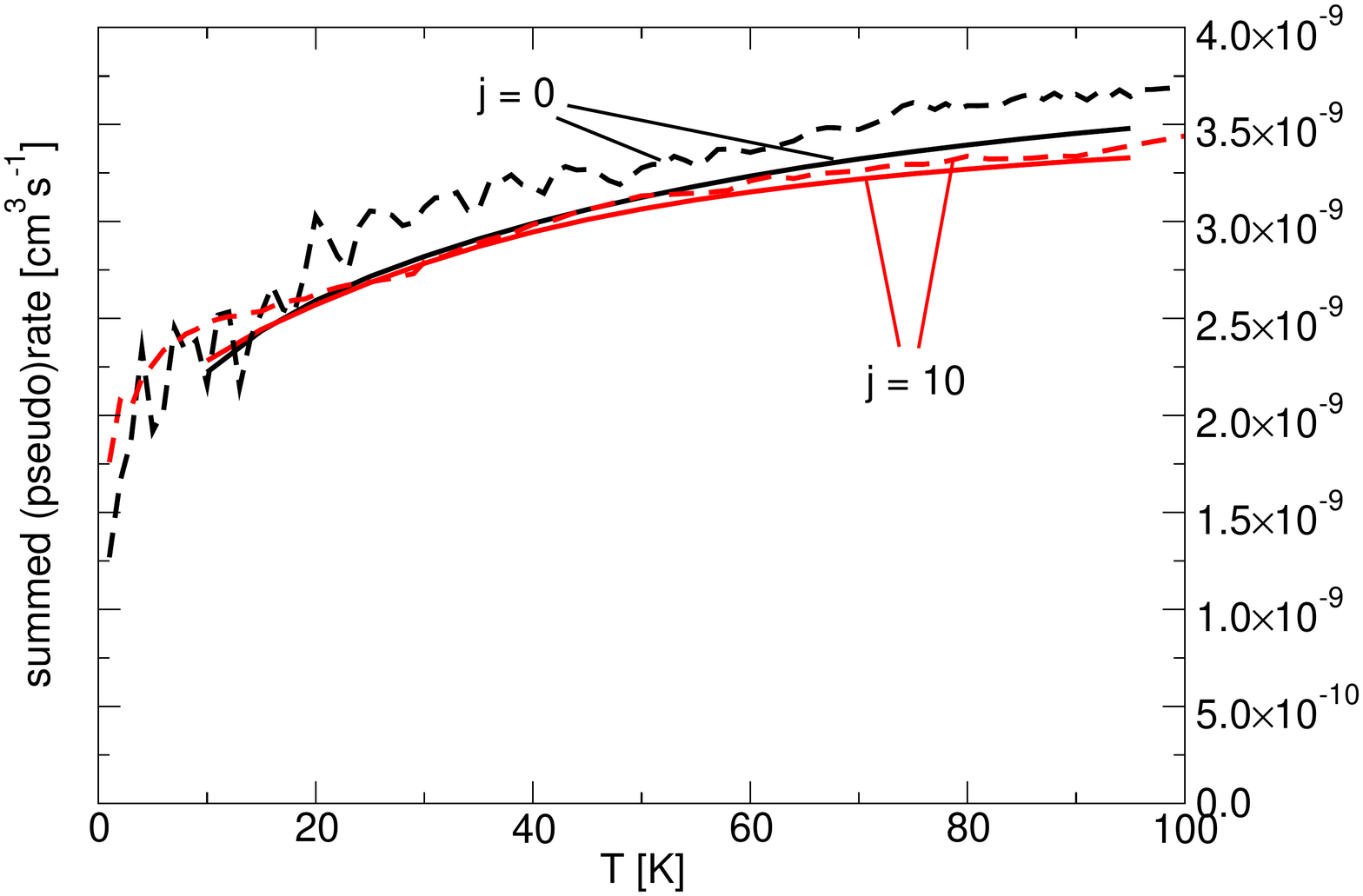}\\
      \includegraphics[width=0.5\textwidth]{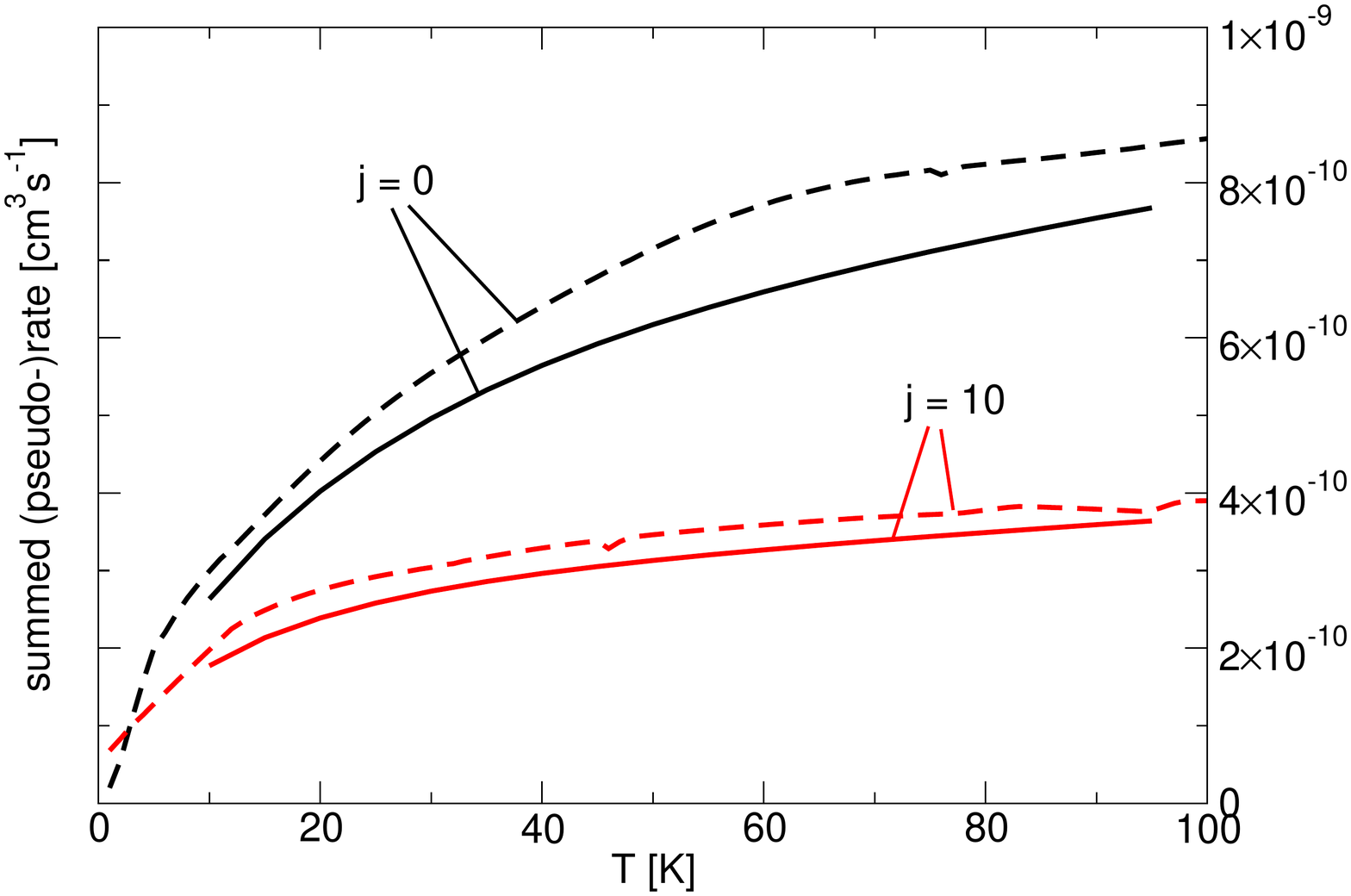}
      &
      \includegraphics[width=0.5\textwidth]{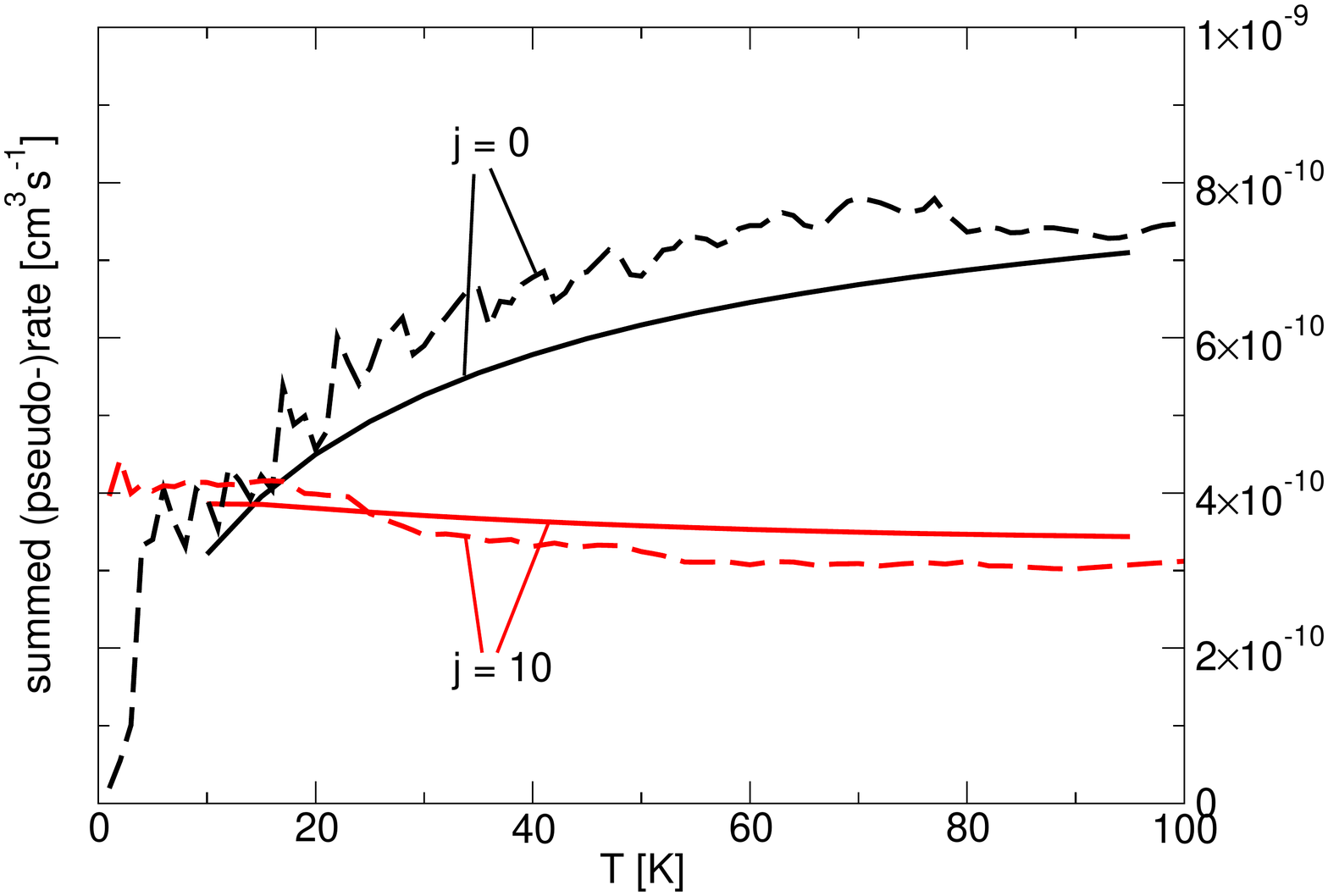}
    \end{tabular}
    \caption{Rotationally summed (pseudo-)rates for the excitation
      (from j=0) and de-excitation (from j=10) processes. Left panels:
      neutral, right panels.: ion. Two different initial states are
      considered by including or excluding the elastic cross
      sections. Pseudo-rates are represented using dashes, while real
      rates are given by full lines. The upper panels are the sums
      including the elastic cross sections, while the lower panels
      plot the sums without the elastic cross sections.}
    \label{fig:rates}
\end{center}
\end{figure}

To assess the reliability of this approximation, we also computed
temperature dependent rates for some transitions and in the range of
10-100K. In figure \ref{fig:rates} we therefore report the
summed-over-final-states Boltzman rates and the pseudo-rates,
$\sum_{j'\le 10} K_{j\to j'}$, as a function of temperature for the
two systems. Three main features are illustrated by the plots: (i) on
all four panels, we see that the pseudo-rates are a very good
approximation to the true Boltzmann-integrated rates. The size and
temperature dependence are largely the same, the main difference being
that the Boltzman integration smooths out the curves and makes the
resonances patterns disappear while it is not the case with the
pseudo-rates: the average precision of this pseudo-rate approximation
can furthermore be estimated to be around 20\%; (ii) outside the
resonance structure shown by the ion, we see that, the global
inelastic behavior of these pseudo-rates is nearly the same for both
molecular partners, the principal difference showing up at low energy,
as was the case for cross sections; (iii) one further difference
between the two systems is to be found in the elastic rates/cross
sections, which turn out to be about twice as big for the ion as for
the neutral. Thus, we can say that the overall flux redistribution
after collisions is dominated by elastic processes in the ionic case,
while for the neutral, the sizes of elastic and inelastic flux
redistributions are nearly equal.

\section{\label{sec:conclusion}Conclusions}
We have computed rotationally inelastic cross sections and pseudo
rates for the lithium dimer in two different electronic states,
treated as pseudo-$^1\Sigma$ molecules, interacting with a helium atom
in the range of energy between 1 and 100 \cm. We found the unexpected
result that, except for the low energy behavior and the resonance
structures present in the ionic case, the inelastic cross sections and
rates are rather close between neutral and ionic partners although the
potential well depths, the repulsive walls and the long range
behaviors are different in the two cases. The explanation comes from
the fact that at these energies, the collisional behavior is dominated
by the geometry effects; in other words, the dynamically accessible
torques at a given energy for a given transition are similar for both
systems. The elastic cross sections and rates are however much more
different, with a factor of 2 in favor of the ion, as one should
expect.

The present calculations therefore help us to shed more light on the
role played by molecular features in low energy inelastic scattering
processes, in the sense that the presence of either neutral or ionized
lithium dimers in the gaseous medium would result, in both cases, in
comparable cooling efficiency for scattering with $^4$He as a buffer
gas. On the other hand, the differences in elastic cross sections
suggest that the ionic partner would yield much larger momentum
transfer cross sections with the same partner gas and would therefore
undergo more rapidly a translational cooling process by sympathetic
collisions (e.g. see \cite{bodo06}).

\section{Acknowledgments}
The financial support of the University of Rome ``La Sapienza''
Research Committee and of the CASPUR Computing Consortium is
gratefully acknowledged. One of us (M.W.) thanks the Department of
Chemistry of ``La Sapienza'' for the award of a Research Fellowship.

\bibliography{paper}

\clearpage

\clearpage

\clearpage

\clearpage

\end{document}